\documentclass[aps,prd,preprint,groupedaddress]{revtex4}

\usepackage{graphicx}

\def\vc#1{\textbf{\textit #1}}
\def\w{\omega}
\def\r2{\sqrt 2}
\def\m#1{{\tilde m}_#1}
\def\sinb{\sin\beta}
\def\cosb{\cos\beta}
\def\tanb{\tan\beta}
\def\sina{\sin\alpha}
\def\cosa{\cos\alpha}
\def\mw#1{m_{\omega_#1}}
\def\PL{\frac{1-\gamma_5}{2}}
\def\PR{\frac{1+\gamma_5}{2}}
\def\PRev#1#2#3{Phys. Rev. {#1}, #2 (#3)}
\def\PRD#1#2#3{Phys. Rev. {\bf D #1}, #2 (#3)}
\def\NPB#1#2#3{Nucl. Phys. {\bf B #1}, #2 (#3)}
\def\PTP#1#2#3{Prog. Theor. Phys. {\bf #1}, #2 (#3)}

\def\PLB#1#2#3{Phys. Lett. {\bf B #1}, #2 (#3)}
\def\PRL#1#2#3{Phys. Rev. Lett. {\bf #1}, #2 (#3)}
\def\PRep#1#2#3{Phys. Rep. {\bf #1}, #2 (#3)}

\begin{document}

\preprint{
OCHA-PP-311
}

\title{
Coexistence of CP eigenstates in Higgs boson decay
}


\author{
Noriyuki Oshimo
}
\affiliation{
Department of Physics,
Ochanomizu University,
Tokyo, 112-8610, Japan
}


\date{\today}

\begin{abstract}

     The supersymmetric extension of the standard model contains an intrinsic source 
 of CP violation mediated by the charginos.  
 As an effect, both CP-even and CP-odd final states could be observed in the Higgs boson 
 decay into two photons whose evidences were reported recently.  

\end{abstract}


\maketitle


     Recently ATLAS \cite{atlas} and CMS \cite{cms} collaborations at LHC reported 
evidences of a new neutral particle with spin 0 and mass around 125 GeV.  
If this particle is the Higgs boson of the standard model (SM), as would be expected, 
the magnitude of mass seems to be rather small.  
Since the Higgs boson mass is constrained only from unitarity of the S matrix 
within the framework of the SM, it is allowed to be any value smaller than about 1 TeV \cite{dicus}.  
On the other hand, if the SM is extended to respect supersymmetry, the Lagrangian is 
more constrained, leading to a much smaller upper bound of about 150 GeV 
on the mass for one of the Higgs bosons \cite{kane}.  
The observed mass value might strengthen plausibility for 
the supersymmetric extension of the SM.  
Furthermore, suggestive information is provided concerning details of the model.  
The mass of the lightest Higgs boson receives large contributions from
radiative corrections mediated by the $t$ quark and $t$ squarks \cite{okada}.  
In order for the Higgs boson to have the observed mass, 
the $t$ squarks are required to be heavier than 1 TeV.  
If supersymmetry soft-breaking masses of the squarks do not depend on flavor, 
then all the squarks should have similar large masses.  

     Large masses for the squarks of the supersymmetric model could be required 
also from the viewpoint of CP invariance, {\it i.e.}, non-observation of 
the electric dipole moment of the neutron \cite{oshimo}.  
Assuming that complex phases involved in the Lagrangian are not suppressed accidentally, 
the squarks are predicted to be heavier than 1 TeV, while the charginos and 
neutralinos can be of order of 100 GeV.  
If the Higgs boson has been found at LHC, it may be implied that the squark masses are 
large and the CP-violating phases are unsuppressed.  
Then, CP violation could well be observed in the processes which are 
mediated by the charginos or neutralinos.  
 
     In this note, based on the minimal supersymmetric extension of the SM,  
we show that CP violation could be revealed in the Higgs boson decay into two photons.  
This decay process is generated at one-loop level, to which the charginos contribute.  
Owing to the complex mass matrix, the interactions of the charginos do not 
conserve CP invariance.  
Consequently, although the lightest neutral Higgs boson is even under CP transformation, 
this CP eigenstate is not maintained in the decaying process.    
The CP-odd final state is yielded at a sizable probability, 
which may be detected by examining the polarization planes of the photons.  

     In the minimal supersymmetric model there are two SU(2)-doublet Higgs 
superfields $H_1$ and $H_2$ with hypercharges $-1/2$ and $1/2$, respectively, 
which contain bosons and fermions.  
Neutral Higgs bosons $\phi_1^0$, $\phi_2^0$ interact with charged 
Higgs fermions $\psi_1^-$, $\psi_2^+$ and SU(2) gauge fermions $\lambda^\pm$ as 
\begin{eqnarray}
   \cal L &=& ig{\phi_1^0}^*\overline{\lambda^-}\PL\psi_1^-
       + ig{\phi_2^0}^*\overline{\lambda^+}\PL\psi_2^+ + \rm{H.c.},
\end{eqnarray}
with $g$ denoting the coupling constant of SU(2) gauge interaction.  
The charginos $\w_i$ $(i=1,2)$ are mass eigenstates of these fermions, 
whose  mass terms are  given by 
\begin{eqnarray}
   \cal L &=& -\left(
      \matrix{\overline{\lambda^-}  & \overline{(i\psi_2^+)^c} }
      \right)M_\w\PL\left(
      \matrix{\lambda^- \cr
                        i\psi_1^-}
      \right) + \rm{H.c.},
      \\
       M_\w &=& \left(
      \matrix{           \m2     & -\frac{1}{\r2}gv_1 \cr
               -\frac{1}{\r2}gv_2 & \mu                   }        
           \right).    
\label{chmass}  
\end{eqnarray}
Here, $\m2$ and $\mu$ stand for the mass parameters originating, 
respectively, from the supersymmetry soft-breaking terms of the SU(2) gauge fermions and 
from the bilinear term of the Higgs superfields in the superpotential.  
These mass parameters have in general complex values, and both complex phases 
cannot be eliminated by redefining the particle fields.  
This is one of the origins of CP violation intrinsic in the supersymmetric SM.  
Without loss of generality, we can take $\mu$ complex 
as $\mu=|\mu|\exp(i\theta)$ and $\m2$ real and positive.  
The vacuum expectation values of the Higgs bosons expressed by $v_1$ and $v_2$ 
are real and positive, with the ratio $v_2/v_1$ being denoted by $\tanb$.   
The mass matrix is diagonalized to give mass eigenstates as
\begin{eqnarray}
      U_R^\dagger M_\w U_L &=& {\rm diag}(\mw1, \mw2),  \quad\left(\mw1 <\mw2\right),   
\end{eqnarray}
where $U_R$ and $U_L$ are unitary matrices.    

     The neutral Higgs bosons yield three physical mass eigenstates, among 
which two states are even and one is odd under CP transformation.  
One of the even eigenstates has the lightest mass of the three.  
Their mass terms are written as 
\begin{eqnarray}
   \cal L &=& -\frac{1}{2}\left(
      \matrix{{\rm Re}(\phi_1^0)  &  {\rm Re}(\phi_2^0)}
      \right)M_H^2\left(
      \matrix{{\rm Re}(\phi_1^0) \cr
                        {\rm Re}(\phi_2^0)}
      \right), 
\end{eqnarray}
and the mass-squared matrix is diagonalized by the orthogonal matrix,      
\begin{eqnarray}  
    U^T M_H^2U &=& {\rm diag}(m_{H1}^2, m_{H2}^2), \quad\left(m_{H1}^2<m_{H2}^2\right),  \\
       U &=& \left(
      \matrix{           -\sina     & \cosa \cr
               \cosa & \sina                  }        
           \right).  
\label{mixing}     
\end{eqnarray}
The mass-squared matrix is given, at tree level, by 
\begin{eqnarray}      
    M_H^2 &=& \left(
     \matrix{      M_Z^2\cos^2\beta +|M_{12}^2|\tanb     & -\frac{1}{2}M_Z^2\sin 2\beta -|M_{12}^2| \cr
         -\frac{1}{2}M_Z^2\sin 2\beta-|M_{12}^2|     &    M_Z^2\sin^2\beta +|M_{12}^2|\cot\beta  }       
           \right), 
     \label{hbmass}  
\end{eqnarray}
where $M_Z$ and $M_{12}^2$ denote respectively the $Z$ boson mass and 
the mass-squared parameter of a supersymmetry soft-breaking term.   
In addition, this matrix receives significant contributions from radiative corrections, 
which depend on various unknown parameters \cite{carena}.  
We therefore take the angle $\alpha$ of the orthogonal matrix for a parameter.  
Radiative corrections could also mix CP eigenstates for the Higgs bosons 
mainly through the interactions with the $t$ squarks, \cite{pilaftsis}.  
However, the large squark masses make these corrections small.  
Moreover, these mixings depend on another source of CP violation, different from 
the mass parameters $\m2$ and $\mu$, originating in the $t$-squark mass-squared matrix. 
Possible CP-violating mixings for the Higgs boson mass eigenstates have been thus neglected.  

     The decay of the lightest Higgs boson into two photons $H_1^0\to\gamma\gamma$ is 
induced at one-loop level by various supersymmetric particles and 
the charged Higgs boson \cite{gunion}, as well as by the particles of the SM \cite{resnick}.  
Assuming the squarks, sleptons, and charged Higgs boson are enough heavy, 
non-negligible contributions are made by the charginos besides the SM particles.  
The relevant interaction Lagrangian concerning the charginos is given by  
\begin{eqnarray}
\cal L &=& g\overline{\w_i}\left(C_{Li}\PL+C_{R_i}\PR\right)\w_i H_1^0
          -e\overline\w_i\gamma^\mu\w_i A_\mu, 
 \label{interaction}\\
   C_{Li} &=& C_{Ri}^* = \frac{1}{\r2}\left(-\sina U_{R1i}^*U_{L2i}+\cosa U_{R2i}^*U_{L1i}\right), 
\end{eqnarray}
where $A_\mu$ stands for the photon field with $e$ being the electric charge.     
Owing to the complex mass matrix for the charginos in Eq. (\ref{chmass}), 
the coefficient $C_L$ has a complex phase, leading to violation of CP invariance.  

     In the decay of the Higgs boson at the rest frame, the helicities of 
two photons are the same, both $h=+1$ or both $h=-1$.  
With $u(\pm,\vc p)$ denoting one photon state with helicity $\pm 1$ and momentum $\vc p$, 
the final state is written as $u(+,\vc p)u(+,-\vc p)$ or $u(-,\vc p)u(-,-\vc p)$.  
These two states are transformed to each other by CP operation, 
so that the eigenstates for CP-even and CP-odd are given respectively by 
\begin{eqnarray}
       f_{\rm even}&=&\frac{1}{\sqrt{2}}\left[u(+,\vc p)u(+,-\vc p)+u(-,\vc p)u(-,-\vc p)\right],  
       \label{cpeven}  \\
       f_{\rm odd}&=&\frac{1}{\sqrt{2}}\left[u(+,\vc p)u(+,-\vc p)-u(-,\vc p)u(-,-\vc p)\right].  
 \label{cpodd}
\end{eqnarray}
Although the lightest Higgs boson is CP-even, both of these final states could 
appear by CP violation.  
The decay widths for the CP eigenstates $ f_{\rm even}$ and  $f_{\rm odd}$ are  given by 
\begin{eqnarray}
    \Gamma_{\rm even} &=& \frac{g^2e^4}{256\pi^5}m_{H1}
    \left|\sum_{i=1}^2 \frac{C_{Li}+C_{Ri}}{2}I(r_{\w i})-\frac{2\cosa}{3\sinb}\frac{m_t}{M_W}I(r_t)
                                        +\sin(\beta-\alpha)K(r_W)\right|^2, \\
   \Gamma_{\rm odd} &=& \frac{g^2e^4}{256\pi^5}m_{H1}
    \left|\sum_{i=1}^2 \frac{-C_{Li}+C_{Ri}}{2}J(r_{\w i})\right|^2, \\
    & & r_{\w i}=\frac{m_{H1}}{m_{\w i}}, \quad r_t=\frac{m_{H1}}{m_t}, 
    \quad  r_W=\frac{m_{H1}}{M_W}, 
\end{eqnarray}
with $m_t$ and $M_W$ being respectively the masses of the $t$ quark and $W$ boson.   
The functions are defined by 
\begin{eqnarray}
 I(r) &=& \frac{2}{r}\left[1-\left(\frac{4}{r^2}-1\right)\left(\arcsin\frac{r}{2}\right)^2\right], \\
 J(r) &=& \frac{2}{r}\left(\arcsin\frac{r}{2}\right)^2, \\
 K(r) &=& \frac{r}{4} + \frac{3}{2r}\left[1-\left(\frac{4}{r^2}-2\right)\left(\arcsin\frac{r}{2}\right)^2\right].  
\end{eqnarray}
Owing to CP-violating interactions in Eq. (\ref{interaction}), the charginos contribute to both 
the widths $\Gamma_{\rm even}$ and $\Gamma_{\rm odd}$.   
The SM particles make contributions only to the CP-even width, among which 
the largest and second largest ones arise from the $W$ boson and $t$ quark 
interactions, respectively.  
The other contributions have been neglected.  
The $t$ quark contribution receives QCD corrections, which however are small \cite{zerwas}.  
The Higgs boson decays dominantly into a pair of $b$ and $\bar b$ quarks.  
Its width is given by  
\begin{equation}
    \Gamma_{\rm bb} = \frac{3g^2}{32\pi}m_{H1}
   \left(\frac{\sina}{\cosb}\frac{m_b}{M_W}\right)^2\left(1-\frac{4m_b^2}{m_{H1}^2}\right)^{3/2}, 
\end{equation}
where $m_b$ denotes the $b$ quark mass.  
Although this width receives non-negligible contributions from radiative corrections \cite{kniehl}, 
the tree level formula would be sufficient for our rough estimation.  

\begin{figure}

\includegraphics{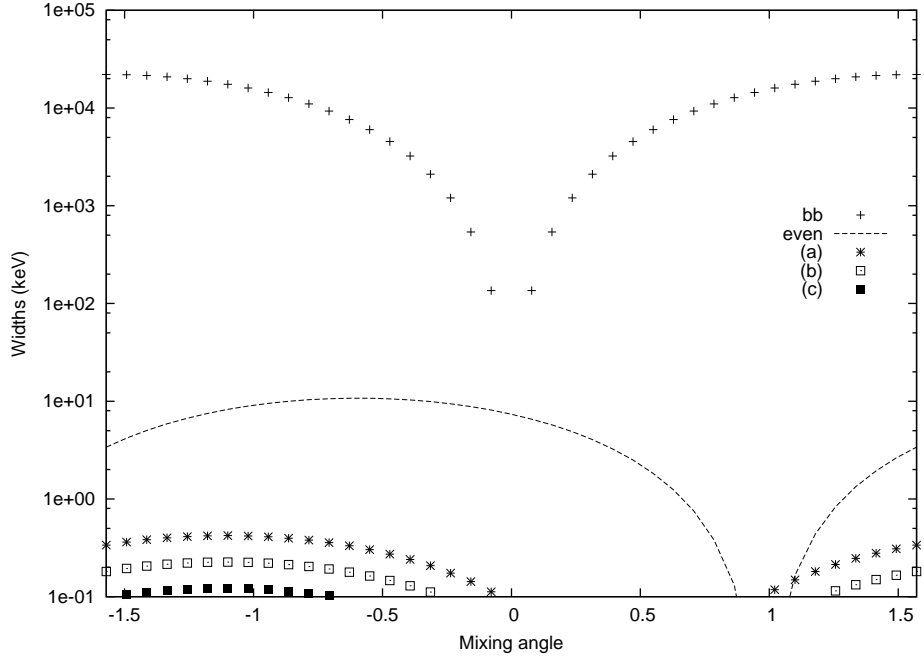}%

(i) $\tanb=2$

\includegraphics{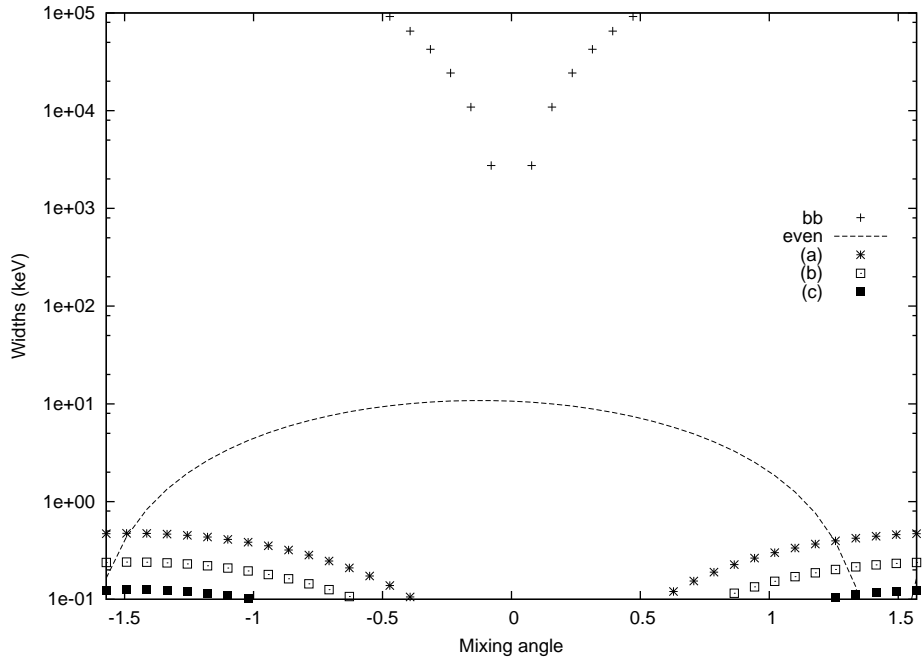}%

(ii) $\tanb=10$

\caption{
Decay width of $H_1^0\to\gamma\gamma$ for the CP-odd final state in cases (a), (b), and (c), 
together with the width for the CP-even final state.  The width of $H_1^0\to \bar bb$ is also depicted.   
\label{widths}
}
\end{figure}

     In Fig. \ref{widths} the widths of the decay $H_1^0\to\gamma\gamma$ 
are shown as functions of the mixing angle $\alpha$ in Eq. (\ref{mixing}), 
together with the width of the dominant decay mode $H_1^0\to\bar bb$ for 
estimating the branching ratios.  
For definiteness, the mass of the lightest Higgs boson $H_1^0$ is taken as $m_{H1}=125$ GeV.  
For the parameters $\m2$ and $|\mu|$ we take three sets of values listed 
in Table \ref{parameters}, with $\tanb=2$ and $\tanb=10$, where
the resultant two chargino masses are given in parentheses.  
The complex phase of $\mu$ is set at $\theta=\pi/2$.  
The dashed line depicts the width $\Gamma_{\rm even}$ in case (b), though 
the other cases lead to almost the same curves.  
The lower three kinds of points depict the width $\Gamma_{\rm odd}$ for the three sets of 
the parameter values.  
The width $\Gamma_{\rm bb}$ is drawn by the upper points.   

     The width for the CP-odd state is larger than 0.1 keV for wide ranges of 
the parameters $\alpha$ and $\tanb$, provided that the charginos are not 
much heavier than 100 GeV.  
Since the CP-even state has the width around 10 keV or smaller, 
in the two photon decay the CP-odd state could appear 
at a rate larger than 1 percent and even around 10 percent.  
Allowed parameter ranges, however, may be constrained by the experimental result for 
the two photon production rate, which shows roughly consistency with the SM.  
This result would suggest that the Higgs boson branching ratio for the two photon decay 
is not different much from the SM value of a few times $10^{-3}$.  
If this constraint is imposed, the parameter region with $\tanb$ around 10 or larger is widely 
excluded unless the magnitude of $\alpha$ and thus the width $\Gamma_{\rm bb}$ are small.  
On the other hand, in the region $\tanb<10$ there exists a sizable range of the parameter $\alpha$ 
which does not cause large deviation from the SM branching ratio.  
In this range the CP-odd state can amount to a few percent of the two photon decay.   
For example, if the parameters have the values $\tanb=2$ and $\alpha=-0.5$, 
the widths satisfy the ratio $\Gamma_{\rm even}/\Gamma_{\rm bb}\simeq2.1\times 10^{-3}$ 
which is approximately the same as the SM value.  
The parameter set (a) in Table \ref{parameters} then  
leads to the widths $\Gamma_{\rm odd}=0.28$ keV and $\Gamma_{\rm even}=10$ keV.  
Smaller magnitudes for $\m2$ and $|\mu|$ make the rate of CP-odd state larger, 
though there is not much room for the parameter values if the experimental 
lower bound on the lighter chargino mass by LEP \cite{pdg} is taken into account.  
Assuming that the production cross section of the Higgs boson is 30 pb 
at the collision energy $\sqrt{s}=14$ TeV \cite{spira}, integrated luminosity 
of 100 fb$^{-1}$ yields $3\times10^6$ Higgs bosons.  
If the branching ratio of the two photon decay is around the SM value,  
CP-odd events of order of 100 are expected from 
the ratio $\Gamma_{\rm odd}/\Gamma_{\rm even}=0.03$.  
It should be noted that there is a parameter region where the branching ratio of 
the two photon decay is larger than the SM value.  
Reported excess of the branching ratio from the SM may be understood through 
the effects of radiative corrections to the mass-squared matrix of 
the Higgs bosons in Eq. (\ref{hbmass}).  

     The two final states of CP-even and CP-odd could be distinguished from each other by  
observing photon polarizations.  
Taking the photon momenta parallel to the $z$ axis at the Higgs boson rest frame, 
we write one photon states in linear polarization parallel to the $x$ axis and 
to the $y$ axis as $u(x,\vc p)$ and $u(y,\vc p)$, respectively.  
The CP- even and CP-odd final states in Eqs. (\ref{cpeven}) and (\ref{cpodd}) are then expressed by 
\begin{eqnarray}
       f_{\rm even}&=&\frac{1}{\sqrt{2}}\left[u(x,\vc p)u(x,-\vc p)+u(y,\vc p)u(y,-\vc p)\right],  \\
       f_{\rm odd}&=&\frac{i}{\sqrt{2}}\left[-u(x,\vc p)u(y,-\vc p)+u(y,\vc p)u(x,-\vc p)\right].  
\end{eqnarray}
In the CP-even state the polarization plane of one photon is parallel to that of another photon, 
while in the CP-odd state the two planes are perpendicular to each other.  
This difference may be detected by examining the angular distributions of 
the leptons or quarks which the photons internally convert to, 
since these observables correlate with the polarization planes.  
For instance, if the photons convert to two pairs of leptons, the angle between 
the planes formed by these lepton momenta can provide information on the 
angle between the polarization planes \cite{kroll}, as determination of parity 
for the neutral $\pi$ meson made use of \cite{plano}.   
Although detailed study including effects of final state interactions is necessary 
for measurement, if perpendicularity of the polarization planes is observed, 
CP violation will be established, favoring the supersymmetric extension 
of the SM.  

     In conclusion, from the recent experimental findings on the Higgs boson, 
it may be suggested that physics beyond the SM is described by 
the supersymmetric extension and its intrinsic sources of CP violation are unsuppressed.  
One possible effect of this scenario is CP violation in the two photon decay 
of the Higgs boson.  
If the chargino masses are around 100 GeV, the Higgs boson in CP-even state 
decays into two photons in CP-odd state at a sizable rate.  
This CP violation could be observed by measuring the polarization planes of the photons.  

\begin{table}
\caption{
Mass parameter values of three examples and resultant chargino masses (GeV).  
${\rm arg}(\mu)=\pi/2$.
\label{parameters}
}
\begin{ruledtabular}
\begin{tabular}{cccc|cc}
  & $\m2$    & $|\mu|$  & & $\tanb=2$ & $\tanb=10$ \\
\hline
 (a) & 150 & 150 & &(107, 216) & (104, 217) \\
 (b) & 200 & 150 & &(124, 245) & (122, 246) \\ 
 (c) & 200 & 200 & &(153, 264) & (151, 265) \\
\end{tabular}
\end{ruledtabular}
\end{table}

\pagebreak

\end{document}